\documentclass[12pt]{iopart}
\usepackage{mathrsfs} 
\expandafter\let\csname equation*\endcsname\relax
\expandafter\let\csname endequation*\endcsname\relax
\usepackage{amsmath}
\usepackage{amsfonts} 
\usepackage{graphicx}
\usepackage{subcaption}
\usepackage{float}
\usepackage{multirow}
\usepackage[explicit]{titlesec}
\usepackage[most]{tcolorbox}
\usepackage{braket}
\usepackage{soul}

\usepackage{indentfirst}
\usepackage{hyperref}
\hypersetup{
    colorlinks=true,
    linkcolor=blue,
    filecolor=magenta,      
    urlcolor=cyan
    }

\begin{document}

\title[]{Exact solution of a  non-Hermitian $\mathscr{PT}$- symmetric  spin chain}

\author{Pradip Kattel$^{*\dagger}$, Parameshwar R Pasnoori$^{\ddagger\star}$, and  Natan Andrei$^{\dagger}$}

\address{$^\dagger$  Department of Physics, Center for Material Theory, Rutgers University, Piscataway, New Jersey 08854, USA}
\address{$^\ddagger$ Department of Physics,
University of Maryland, College Park, Maryland 20742}
\address{$\star$ Laboratory for Physical Sciences, 8050 Greenmead Dr, College Park, Maryland 20740}
\ead{pradip.kattel@rutgers.edu}
\vspace{10pt}
\begin{indented}
\item[]January 2023
\end{indented}

\begin{abstract}
 We construct the exact solution of a  non-Hermitian $\mathscr{PT}$-symmetric isotropic Heisenberg spin chain with integrable boundary fields. We find that the system exhibits two types of phases named $A$ and $B$. In the $B$ type phase, the $\mathscr{PT}$ symmetry remains unbroken and it comprises of eigenstates with only real energies, whereas the $A$ type phase exhibits both $\mathscr{PT}$ symmetry broken and unbroken sectors, comprising of eigenstates with only complex and real energies respectively. The $\mathscr{PT}$-symmetry broken sector comprises of pairs of eigenstates whose energies are complex conjugates of each other. The existence of two sectors in the $A$ type phase is associated with the exponentially localized bound states at the edges with complex energies which are described by boundary strings. We find that both $A$ and $B$ type phases can be further divided into sub-phases which exhibit different ground states. We also compute the bound state wavefunction in one magnon sector and find that as the imaginary value of the boundary parameter is increased, the exponentially localized wavefunction broadens thereby protruding more into the bulk, which indicates that exponentially localized bound states may not be stabilized for large imaginary values of the boundary parameter.
 
\end{abstract}

%

\section{Introduction}

Quantum dynamics of physical systems is governed by Hamiltonians whose eigenvalues give the energies of the  states in the system. In a conventional isolated quantum system, the Hamiltonian is required to be Hermitian so that its eigenvalues are real and the time evolution is unitary, thereby guaranteeing the conservation of probabilities. This picture of quantum mechanics is quite idealized as no real quantum system exists in isolation. One way to incorporate the effect of the environment in a quantum system is to consider effective non-Hermitian Hamiltonians which naturally arises when the full Hermitian Hamiltonian describing the system and its environment is projected  on the space of the system of interest \cite{breuer2002theory}. These Hamiltonians generally have complex eigenspectra and they describe  out-of-equilibrium systems which may gain or lose energy  to the environment. The imaginary part of the energy is related to the lifetime of a state of the system.

 An important class of non-Hermitian  Hamiltonians, known as space-time reflection symmetric,   or  $\mathscr{PT}$ -symmetric Hamiltonians,  was introduced in a seminal paper by Bender and Boettcher \cite{bender1998real}. 
 These Hamiltonians describe the borderline systems between the open and closed systems in the sense that they are not isolated like the closed systems but the effect of the environment on them is severely constrained as the loss and gain in the system are balanced out such that there is no net loss or gain in the system. Due to this constraint, $\mathscr{PT}$-symmetric Hamiltonians can have real energy levels. $\mathscr{PT}$-symmetric quantum mechanics and quantum field theories have found applications in wide range of areas including quantum optics \cite{castaldi2013p,yin2013unidirectional,zyablovsky2014pt,klauck2019observation}, and condensed matter systems \cite{kawabata2018parity,kornich2022signature,bagarello2015non,zhao2017p,turker2018pt}.

 The eigenvalues of such Hamiltonians are either real, corresponding to   $\mathscr{PT}$-symmetric eigenvectors,  or 
 appear as complex conjugate pairs when the corresponding eigenvectors  break the $\mathscr{PT}$ - symmetry. This follows from the anti-linearity  of the $\mathscr{PT}$ - reflection symmetry, which implies that although $[\mathscr{PT},H]=0$,  the Hamiltonian and the $\mathscr{PT}$ operator  may not be simultaneously diagonalizable.  Consider first the eigenvalue equation $H\ket\psi=\lambda\ket\psi$ for a
state $\ket\psi$ that is symmetric, namely $\mathscr{PT}\ket\psi=\ket\psi$, then
$0=[\mathscr{PT},H]\ket{\psi}=(\lambda^*-\lambda)\ket\psi$ implies that the eigenvalue is real. However 
when the symmetry is spontaneously broken, namely $\mathscr{PT}\ket\psi=\ket{\psi'}\neq \ket\psi$, then $0=[\mathscr{PT},H]\ket{\psi}$ implies $H\ket{\psi'}=\lambda^*\ket{\psi'}$ indicating that if there exist a state $\ket{\psi}$ with complex energy $\lambda$, then there is another state $\ket{\psi'}$ with complex conjugate energy  $\lambda$. The converse proof for both of the cases also holds.\footnote{  If the eigenvectors of $\mathscr{PT}-$symmetric Hamiltonians are $\mathscr{PT}-$symmetric \textit{i.e.} the $\mathscr{PT}-$symmetric is not broken, then the eigenspecra are entirely real. But if the eigenvectors of $\mathscr{PT}-$symmetric Hamiltonians are not $\mathscr{PT}-$symmetric \textit{i.e.} the $\mathscr{PT}-$symmetric is spontaneously broken, then the eigenspecra appear as complex conjugate pairs. Moreover, there can also be exceptional points, characterized by
the coalescence of one or multiple pairs of eigenvalues and eigenvectors, where Hamiltonian is not diagonalizable\cite{kato2013perturbation,berry2004physics}.} The phase transitions between the unbroken phase and the spontaneously broken $\mathscr{PT}$-phase have been experimentally realized in different systems \cite{chitsazi2017experimental,sun2014experimental,Schindler2011ObservationOS}.

Computing  analytically the energy levels of the $\mathscr{PT}$-symmetric Hamiltonians is a difficult task. Even simple single-particle quantum mechanics with $\mathscr{PT}$-symmetric potentials like $ix^3$ potential \cite{bender2004extension} or $\mathscr{PT}$- symmetric square well potential \cite{bender2006calculation} can only be solved perturbatively. The proof of the reality of energy levels of certain $\mathscr{PT}$-symmetric quantum mechanics uses the conjecture of correspondence between ordinary differential equation and integrable models \cite{dorey2004reality}. However, general proof of the conjecture is not known. Even though $\mathscr{PT}$-symmetric spin chains have  been studied focusing on various aspects \cite{castro2009spin,korff2008pt,korff2007pt}, and $\mathscr{PT}$-symmetric deformation of integrable models\cite{fring2013pt} and some other $\mathscr{PT}$-symmetric many-body systems \cite{joglekar2010robust,zhu2014pt,giorgi2010spontaneous} have been studied both numerically and analytically, there has been no complete solution showing that all the excited states can have either real energies or they form pairs having complex conjugate energies. {{Some non-interacting exactly solvable many-body models have been analytically studied before \cite{baxter1989simple,fendley2014free}}}.


In this paper, we introduce a many-body $\mathscr{PT}$-symmetric model which is constructed by applying complex magnetic fields to the edges of a Heisenberg spin chain. The Hamiltonian is given by 

\begin{equation}
    \mathcal{H}=\sum_{j=1}^{N-1}\sum_{\alpha=\{x,y,z \}}\sigma_j^\alpha\cdot\sigma_{j+1}^\alpha+h_1\sigma_1^z+h_N\sigma_N^z,
    \label{opnham}
\end{equation}
where $h_1$ and $h_N$ are the magnetic fields acting at the first and last site in the $z$ - direction, $h_1= h-i\gamma, h_N=h+i\gamma$. Equivalently, upon the Jordan-Wigner transformation, we obtain a fermionic version of the Hamiltonian
\begin{equation}
    \mathcal{H}_f=\frac{J}{2}\sum_{i=1}^{N-1} f_{j+1}^\dagger f_j+f_j^\dagger f_{j+1}+J\left(n_j-\frac12\right)\left(n_{j+1}-\frac12\right)+h_1n_1+h_Nn_N,
\end{equation}
which describes a fermionic chain with nearest-neighbor interactions coupled to baths at its edges, as shown in the figure below. This model may be experimentally realized in cold atom systems or optical systems with controlled loss and gain \cite{kreibich2014realizing,takasu2020pt,wang2022pt}.

The exact results presented in this work might be useful for providing the benchmark for various numerical techniques developed for non-Hermitian quantum mechanics\cite{chan2005density,wessels2009numerical} .

\begin{figure}[H]
    \centering
    \subfloat[\centering Spin chain interacting with bath]{{\includegraphics[width=.5\textwidth]{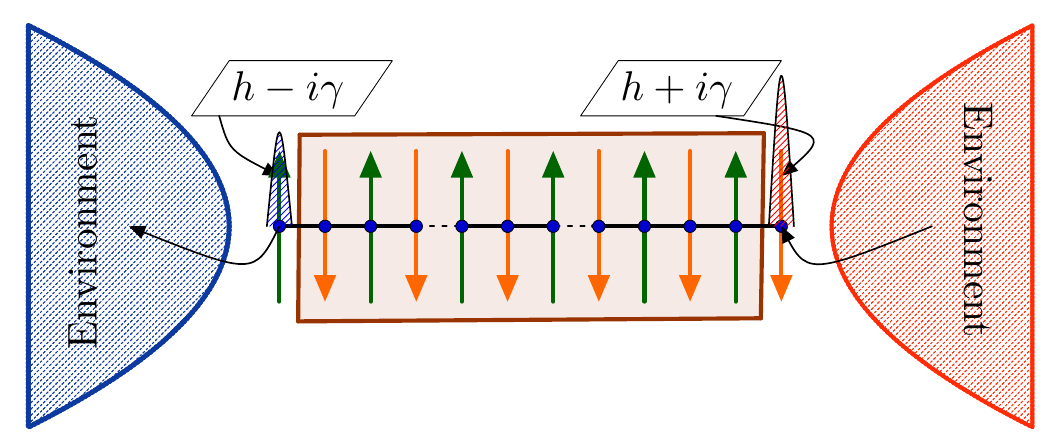} }}%
    \qquad
    \subfloat[\centering Ends of fermion chain interacting with bath]{{\includegraphics[width=.5\textwidth]{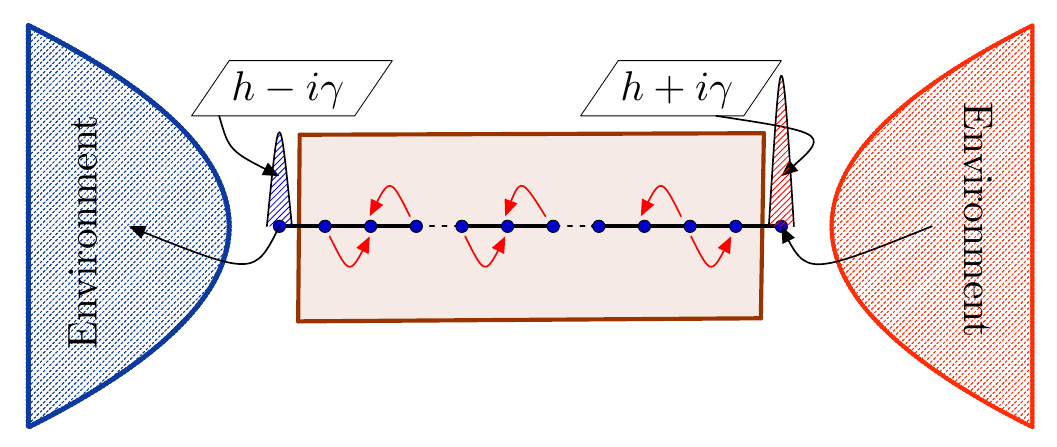} }}%
    \caption{Cartoon showing physical realization of the model.}%
    \label{fig:example}%
\end{figure}

This system with real values of the boundary fields was solved using Bethe Ansatz \cite{alcaraz1987surface,sklyanin1988boundary,grisaru1995direct} and recently the complete phase diagram and the structure of the Hilbert space were obtained \cite{XXXmag}.

In this work we consider the case where $\gamma\neq 0$ and solve the system exactly using Bethe Ansatz. We find that the system exhibits several phases which can be broadly classified into $A$ type and $B$ type phases. In the $B$ type phases the $\mathscr{PT}$ - symmetry remains unbroken, whereas $A$ type phases exhibit both $\mathscr{PT}$ - symmetry broken and unbroken sectors. The existence of two sectors in the $A$ phases is related to the existence of bound state solutions at the edges with complex energies, where the energy of the bound state at the left edge is a complex conjugate of the energy of the bound state at the right edge. Furthermore, we find that the ground state exhibited by the system depends on the orientation of the boundary fields, hence, both the $A$ and $B$ type phases can be further divided into sub-phases depending on the ground state exhibited by the system. Notice that $\gamma\to 0$  corresponds to both the edge fields taking equal and real values, which corresponds to a special case of \cite{XXXmag}. In this limit, we recover the results in \cite{XXXmag} corresponding to this special case of equal fields: We find that when the values of these fields are greater than the critical value $h=2J$, there are stable edge modes, and if the fields are below the critical value, there are exist no edge modes. In addition to this, the energies of the bound states of \cite{XXXmag} are reproduced in this limit where $\gamma\to 0$.

The paper is organized as follows: We summarize the Bethe Ansatz results in the section \ref{sec:summary} and present the properties of the model  in the section \ref{sec:model}. We present the details of the solution in sections \ref{sec:BAequations} \ref{sec:BAsolution} and \ref{sec:WF}, and finally discuss the results in section \ref{sec:discussion}. 



\section{Summary of the Results}
\label{sec:summary}
In this section, we summarize the results obtained via the Bethe Ansatz, relegating the details of the solution to the later sections. There exist four phases that correspond to different ranges of the parameter $\xi=\Re(1/h_1)=\Re(1/h_2)$. The phases $B_1$ and $B_2$ correspond to the ranges $\xi>\frac{1}{2}$ and $\xi<-\frac{1}{2}$ respectively, whereas the phases $A_1$ and $A_2$ correspond to the ranges $0<\xi<\frac{1}{2}$ and $-\frac{1}{2}<\xi<0$ respectively. Below we summarize the results in each of these phases separately for odd and even numbers of sites chain

\subsection{Ground state}
\subsubsection{Odd number of sites}
In phases $B_1$ and $A_1$, the ground state has a total spin $S^z=-\frac{1}{2}$ whereas in phases $B_2$ and $A_2$, the ground state has total spin $S^z=\frac{1}{2}$. In both cases, the total spin in the ground state corresponds to a static spin distribution. The energy of the ground state in each of these phases is real and is equal to $E_0$, whose form is given by \eqref{engeqnreim}.

\subsubsection{Even number of sites}
In phases $B_1$ and $A_1$, the ground state is two-fold degenerate and contains a spinon on top of the static spin distribution with spin $S^z=-\frac{1}{2}$ corresponding to the ground state of an odd number of sites chain. Hence the ground state has a total spin $S^z=-1,0$, corresponding to the spin orientation of the spinon pointing in the negative and positive $z$ directions respectively.

In phases $B_2$ and $A_2$, the ground state is two-fold degenerate and contains a spinon on top of the static spin distribution with spin $S^z=+\frac{1}{2}$ corresponding to the ground state of an odd number of sites chain. Hence, the ground state has a total spin $S^z=1,0$, corresponding to the spin orientation of the spinon pointing in the negative and positive $z$ directions respectively.

\subsection{Excitations}
\subsubsection{$B$ phases}
In the $B$ phases there exist no bound states at both edges. One can build up excitations on top of the ground state by adding an even number of spinons, bulk strings, and quartets, where the energy of all these excitations is real. 

\subsubsection{ $A$ phases}
In the $A$ phases, similar to the $B$ phases, one can build up excitations on top of the ground state and obtain excitations which have real energies. 

There exist two bound state solutions corresponding to the left and right edges. The bound state energy of the left and right boundaries are complex conjugates of each other and are given by the expressions \eqref{boundenrealleft},\eqref{boundenimagleft}\eqref{boundenrealright},\eqref{boundenimagright}. In order to add a bound state to either the left or the right edge, one needs to add a spinon, whose energy is given by \eqref{energyspinon}. The lowest energy of the spinon corresponds to rapidity $\theta\rightarrow\infty$.  Hence the state which contains a bound state at the left or the right edge has complex energy and forms a continuous branch parameterized by the rapidity $\theta$ of the spinon. 

One can build excitations on top of each of these states containing one bound state, by adding an even number of spinons, wide boundary strings, bulk strings, and quartets \cite{XXXmag}. All these excitations have complex energies due to the presence of a single bound state either at the left or right edge. 

One can add both the bound states to the ground state which does not require one to add a spinon. Since the bound states have energies that are complex conjugates of each other, this state has real energy. One can construct excitations on top of this state by adding an even number of spinons, bulk strings, wide boundary strings and quartets. All of the excited states built this way have real energy.

Hence, in the $A$ phases, there exist excited states built on top of the state with no bound states or on top of the state which contains two bound states, and obtain excitations whose energies are real. These states correspond to $\mathscr{PT}-$symmetry unbroken sector. Likewise, there exist excited states built on top of the states which contain a bound state either at the left or the right edge, and one obtains excitations which have complex energies, and hence they correspond to $\mathscr{PT}-$symmetry broken sector. The results of the odd and even number of sites chain are summarized in the tables (\ref{tab:odd}),(\ref{tab:even}) respectively.

\begin{table}[H]
\caption{Energies and $\mathscr{PT}-$symmetry of the ground state and the lowest energy states containing bound states for an odd number of sites for various ranges of boundary parameter $\xi=\frac{h}{\gamma ^2+h^2}$. The $B$ phases only consist of symmetry unbroken states with real energy eigenvalues whereas the $A$ phases consist of both broken and unbroken $\mathscr{PT}-$ symmetric states.}
\label{tab:odd}
\begin{tabular}{|l|l|l|l|l|l|}
\hline
Phase &
  Parametric Range &
  State &
  Spin &
  Energy &
  $\mathscr{PT-}$symmetry \\ \hline
$B_2$ &
  $\xi<-\frac{1}{2}$ &
  $\left|\frac{1}{2}\right\rangle$ &
  $\frac{1}{2}$ &
  $E_0$ &
  Unbroken \\ \hline
\multirow{2}{*}{$A_2$} &
  \multirow{2}{*}{$-\frac{1}{2}<\xi<0$} &
  \begin{tabular}[c]{@{}l@{}}$\left|\frac{1}{2}\right\rangle$\\ $\left|-\frac{1}{2}\right\rangle_{L/R}$\end{tabular} &
  \begin{tabular}[c]{@{}l@{}}$\frac{1}{2}$\\ $-\frac{1}{2}$\end{tabular} &
  \begin{tabular}[c]{@{}l@{}}$E_0$\\ $E_0+m+m^*$\end{tabular} &
  Unbroken \\ \cline{3-6} 
 &
   &
  \begin{tabular}[c]{@{}l@{}}$\left|\pm\frac{1}{2}\right\rangle_{\theta,L}$\\ $\left|\pm\frac{1}{2}\right\rangle_{\theta,R}$\end{tabular} &
  \begin{tabular}[c]{@{}l@{}}$\pm\frac{1}{2}$\\ $\pm\frac{1}{2}$\end{tabular} &
  \begin{tabular}[c]{@{}l@{}}$E_0+E_\theta+m$\\ $E_0+E_\theta+m^*$\end{tabular} &
  Broken \\ \hline
\multirow{2}{*}{$A_1$} &
  \multirow{2}{*}{$0<\xi<\frac{1}{2}$} &
  \begin{tabular}[c]{@{}l@{}}$\left|-\frac{1}{2}\right\rangle$\\ $\left|\frac{1}{2}\right\rangle_{L/R}$\end{tabular} &
  \begin{tabular}[c]{@{}l@{}}$-\frac{1}{2}$\\ $\frac{1}{2}$\end{tabular} &
  \begin{tabular}[c]{@{}l@{}}$E_0$\\ $E_0+m+m^*$\end{tabular} &
  Unbroken \\ \cline{3-6} 
 &
   &
  \begin{tabular}[c]{@{}l@{}}$\left|\pm\frac{1}{2}\right\rangle_{\theta,L}$\\ $\left|\pm\frac{1}{2}\right\rangle_{\theta,R}$\end{tabular} &
  \begin{tabular}[c]{@{}l@{}}$\pm\frac{1}{2}$\\ $\pm\frac{1}{2}$\end{tabular} &
  \begin{tabular}[c]{@{}l@{}}$E_0+E_\theta+m$\\ $E_0+E_\theta+m^*$\end{tabular} &
  Broken \\ \hline
$B_1$ &
  $\xi>\frac{1}{2}$ &
  $\left|-\frac{1}{2}\right\rangle$ &
  $-\frac{1}{2}$ &
  $E_0$ &
  Unbroken \\ \hline
\end{tabular}
\end{table}

\begin{table}[H]
\caption{Energies and $\mathscr{PT}-$symmetry of the ground states and the lowest energy states with bound states for even number of sites for various ranges of boundary parameter $\xi=\frac{h}{\gamma ^2+h^2}$. Similar to the odd number of sites chain, the $B$ phases only consist of symmetry unbroken states with real energy eigenvalues whereas the $A$ phases consist of both broken and unbroken $\mathscr{PT}-$ symmetric states.}
\label{tab:even}
\begin{tabular}{|l|l|l|l|l|l|}
\hline
Phase &
  Parametric Range &
  State &
  Spin &
  Energy &
  $\mathscr{PT-}$symmetry \\ \hline
$B_2$ &
  $\xi<-\frac{1}{2}$ &
  \begin{tabular}[c]{@{}l@{}}$\ket{0}_\theta$\\ $\ket{1}_\theta$\end{tabular} &
  \begin{tabular}[c]{@{}l@{}}$0$\\ $1$\end{tabular} &
  \begin{tabular}[c]{@{}l@{}}$E_0+E_\theta$\\ $E_0+E_\theta$\end{tabular} &
  Unbroken \\ \hline
\multirow{2}{*}{$A_2$} &
  \multirow{2}{*}{$-\frac{1}{2}<\xi<0$} &
  \begin{tabular}[c]{@{}l@{}}$\ket{1}_\theta$\\ $\ket{0}_\theta$\\ $\ket{1}_{\theta,L,R}$\\ $\ket{0}_{\theta,L,R}$\end{tabular} &
  \begin{tabular}[c]{@{}l@{}}$1$\\ $0$\\ $1$\\ $0$\end{tabular} &
  \begin{tabular}[c]{@{}l@{}}$E_0+E_\theta$\\ $E_0+E_\theta$\\ $E_0+E_\theta+m+m^*$\\ $E_0+E_\theta+m^*+m$\end{tabular} &
  Unbroken \\ \cline{3-6} 
 &
   &
  \begin{tabular}[c]{@{}l@{}}$\ket{0}_L$\\ $\ket{0}_R$\end{tabular} &
  \begin{tabular}[c]{@{}l@{}}$0$\\ $0$\end{tabular} &
  \begin{tabular}[c]{@{}l@{}}$E_0+m$\\ $E_0+m^*$\end{tabular} &
  Broken \\ \hline
\multirow{2}{*}{$A_1$} &
  \multirow{2}{*}{$0<\xi<\frac{1}{2}$} &
  \begin{tabular}[c]{@{}l@{}}$\ket{-1}_\theta$\\ $\ket{0}_\theta$\\ $\ket{-1}_{\theta,L,R}$\\ $\ket{0}_{\theta,L,R}$\end{tabular} &
  \begin{tabular}[c]{@{}l@{}}$-1$\\ $0$\\ $-1$\\ $0$\end{tabular} &
  \begin{tabular}[c]{@{}l@{}}$E_0+E_\theta$\\ $E_0+E_\theta$\\ $E_0+E_\theta+m+m^*$\\ $E_0+E_\theta+m^*+m$\end{tabular} &
  Unbroken \\ \cline{3-6} 
 &
   &
  \begin{tabular}[c]{@{}l@{}}$\ket{0}_L$\\ $\ket{0}_R$\end{tabular} &
  \begin{tabular}[c]{@{}l@{}}$0$\\ $0$\end{tabular} &
  \begin{tabular}[c]{@{}l@{}}$E_0+m$\\ $E_0+m^*$\end{tabular} &
  Broken \\ \hline
$B_1$ &
  $\xi>\frac{1}{2}$ &
  \begin{tabular}[c]{@{}l@{}}$\ket{-1}_\theta$\\ $\ket{0}_\theta$\end{tabular} &
  \begin{tabular}[c]{@{}l@{}}$-1$\\ $0$\end{tabular} &
  \begin{tabular}[c]{@{}l@{}}$E_0+E_\theta$\\ $E_0+E_\theta$\end{tabular} &
  Unbroken \\ \hline
\end{tabular}
\end{table}


\section{Many body $\mathscr{PT}$- invariant Hamiltonian}
\label{sec:model}
The isotropic Heisenberg $XXX_{\frac{1}{2}}$ Hamiltonian  is a linear endomorphism in the product space $\displaystyle\otimes_{j=1}^N h_j$ where the local Hilbert space of each quantum spin variable is the two-dimensional complex vector space $h_n=\mathbb{C}^2$
\begin{equation}
\mathcal{H}=\sum_{j=1}^{N}\sum_{\alpha=\{x,y,z \}}\sigma_j^\alpha\cdot\sigma_{j+1}^\alpha,
\end{equation}
where $\sigma_j^\alpha$ are the Pauli matrices acting on vector space $h_j$. Imposing the periodic boundary condition, $\sigma_1^\alpha=\sigma_{N+1}^\alpha$, Bethe solved the eigenvalue problem $H\psi=E\psi$ exactly using a method now called \textit{Bethe Ansatz method} \cite{bethetheory}. Alcaraz et al. solved the model with open boundary conditions applying real boundary fields \cite{alcaraz1987surface}. Here we consider the complex deformation as mentioned in Hamiltonian Eq.\eqref{opnham} by allowing the boundary fields to be complex \textit{i.e.} $h_{1/N}\in\mathbb{C}$. Our strategy is to use the Algebraic Bethe Ansatz along with Cherednik-Sklyanin reflection algebra to diagonalize the Hamiltonian \cite{sklyanin1988boundary}. The complex boundary terms lower the $SU(2)$ symmetry to $U(1)$ symmetry $[\mathcal{H},\sum_i S_i^z]=0$ . Thus, the total z-component of the spin $M$ is still a good quantum number for the model. The boundary terms, however, break the $\mathbb{Z}_2$ flip symmetry. The bulk is symmetric under the $\mathbb{Z}_2$ flip, but the entire system with boundary remains invariant only under the following transformation
\begin{equation}
\mathcal{H}\to\prod_{i=1}^N \sigma_i^x\mathcal{H}\sigma_i^x, \quad h_1\to-h_1 \quad \text{and}\quad h_N\to-h_N.
\label{symeqn}
\end{equation}

\section{Bethe Ansatz Equations}
\label{sec:BAequations}
We wish to solve the eigenvalue problem for the Hamiltonian given by Eq.\eqref{opnham} $\mathcal{H}\in \text{End}\left(\stackrel{N}{\otimes}h_n\right)$  with complex boundary fields using \textit{Algebraic Bethe Ansatz}. 

In $V_a\otimes h_n ~\exists$ a L-matrix
\begin{equation}
L_{a,n}(\lambda)=\left(\lambda-\theta_n\right)I_{a,n}+P_{a,n} ,
\end{equation}
where $V_a=\mathbb{C}^2$ is an auxiliary space and $\theta_n$ is the site-dependent inhomogeneity parameter. The L-matrix satisfies the fundamental commutation relation
\begin{equation}
R_{a,b}(\lambda-\mu)L_{n,a}(\lambda)L_{n,b}(\mu)=L_{n,b}(\mu)L_{n,a}(\lambda)R_{a,b}(\lambda-\mu) ,
\label{RLLtrl}
\end{equation}
where the R-matrix acting as an intertwiner $\in V_a\otimes V_b$  given by
\begin{equation}
R_{a,b}(\lambda)=\lambda I_{a,b}+ P_{a,b}
\label{Rmat}
\end{equation}
is a matrix that solves the Yang-Baxter equation
\begin{equation}
R_{12}(\lambda-\lambda')R_{13}(\lambda)R_{23}(\lambda')=R_{23}(\lambda')R_{13}(\lambda)R_{12}(\lambda-\lambda'),
\end{equation}
and satisfies the unitary condition
\begin{equation}
R_{12}(\lambda)R_{21}(-\lambda)\propto \mathbb{I}.
\end{equation}
To describe the open-boundary condition, apart from the $RLL$ relation Eq.\eqref{RLLtrl}, the R-matrix also has to satisfy Sklyanin’s reflection algebra given by
\begin{equation}
R_{1,2} (\lambda-\mu) K_{1}^{-}(\lambda) R_{2,1}(\lambda+\mu) K_{2}^{-}(\mu)=K_{2}^{-}(\mu) R_{1,2}(\lambda+\mu) K_{1}^{-}(\lambda) R_{2,1}(\lambda-\mu).
\label{refalg}
\end{equation}

The diagonal $K^-$ matrix that satisfies the reflection equation \eqref{refalg} is \cite{sklyanin1988boundary,de1994boundary}

\begin{equation}
K^-(\lambda)=\begin{pmatrix}
\xi_-+\lambda & 0 \\
0 & \xi_--\lambda
\end{pmatrix},
\end{equation}
where $\xi_-$ is a free complex parameter.

There exists a dual reflection algebra given by
\begin{equation}\label{dualrefeq}
R_{1,2}(-\lambda+\mu) K_{1}^{+}(\lambda) R_{2,1}(-\lambda-\mu-2 ) K_{2}^{+}(\mu)=K_{2}^{+}(\mu) R_{1,2}(-\lambda-\mu-2 ) K_{1}^{+}(\lambda) R_{2,1}(-\lambda+\mu).
\end{equation}
The $K^+$ matrix that satisfies the dual reflection equation \eqref{dualrefeq} is \cite{sklyanin1988boundary,de1994boundary}
\begin{equation}
K^+(\lambda)=\begin{pmatrix}
\xi_++\lambda+1 & 0 \\
0 & \xi_+-\lambda-1
\end{pmatrix},
\end{equation}
where $\xi_+$ is a free parameter.
One can define a transfer matrix shown below
\begin{equation}
t(\lambda)=\text{tr}_a\left(K_a^+(\lambda)T_{a}(\lambda) K_{a}^{-}(\lambda) \hat{T}_{a}(\lambda) \right),
\label{transmat}
\end{equation}
where the trace is over the auxiliary space $V_a$ and the operators $T_a(\lambda)$ and $\hat T_a(\lambda)$ are given by
\begin{align}
T_a(\lambda)&=L_{a,N}(\lambda)L_{a,N-1}(\lambda)\cdots L_{a,2}(\lambda)L_{a,1}(\lambda),\\
\hat T_a(\lambda)&=L_{a,1}(\lambda)L_{a,2}(\lambda)\cdots L_{a,N-1}(\lambda)L_{a,N}(\lambda).
\end{align}

It is straightforward to prove the commutative property of the transfer matrix for any two distinct spectral parameters \textit{i.e.} $[t(\lambda_i), t(\lambda_j)] = 0~\forall~ \lambda_i,\lambda_j\in\mathbb{C}$ and its eigenvalues are holomorphic functions of $\lambda\in\mathbb{C}$. Moreover, the commutative property guarantees that the transfer matrix behaves as a generating function for a family of infinite towers of commuting operators, among which is the Hamiltonian. Simultaneous diagonalization of the operators $t(\lambda_i)$ allows us to obtain complete  spectral characteristics of the quantum integrable model.  In this model, Hamiltonian is related to the first derivative of the logarithm of $t(\lambda)$  at $\lambda=0$ and $\theta_i=0~ \forall ~i$ \textit{i.e.}
\begin{align}
\mathcal{H} &=\left.\frac{\partial \ln t(\lambda)}{\partial \lambda}\right|_{\lambda=0,\left\{\theta_{j}=0\right\}}-N \nonumber\\
&=\sum_{j=1}^{N-1}\sum_{\alpha=1}^3 \sigma_{j}^\alpha  \sigma_{j+1}^\alpha+\frac{1}{\xi_-} \sigma_{1}^{z}+\frac{1}{\xi_+}\sigma_{N}^{z},
\label{logdtu}
\end{align}
if one identifies the free complex parameters $\xi_\pm$ as $\xi_-=\frac{1}{h_1}$ and $\xi_+=\frac{1}{h_N}$.

The eigenvalues $\Lambda(\lambda)$ of the transfer matrix Eq.\eqref{transmat} satisfy the Baxter's $T-Q$ relation \cite{baxter1972partition}
\begin{equation}
\begin{aligned}
\Lambda(\lambda)=& \frac{2(\lambda+1)^{2 N+1}}{2 \lambda+1}(\lambda+\xi_-)\left( \lambda+\xi_+\right)\frac{Q(\lambda-1)}{Q(\lambda)} \\
&+\frac{2 \lambda^{2 N+1}}{2\lambda+1}(\lambda+1-\xi_-)\left(\lambda+1-\xi_+\right) \frac{Q(\lambda+1)}{Q(\lambda)} ,
\end{aligned}
\label{TQrel}
\end{equation}
where the Q-function is given by
\begin{equation}
Q(\lambda)=\prod_{\ell=1}^M (\lambda-\lambda_\ell	)(\lambda+\lambda_\ell+1).
\end{equation}
Regularity of the T-Q equation gives the BAEs corresponding to the reference state where all spins are up
\begin{equation}
\left(\frac{\lambda_j+1}{\lambda_j}\right)^{2N} \frac{(\lambda_j+\xi_-)(\lambda_j+\xi_+)}{(\lambda_j+1-\xi_-)(\lambda_j+1-\xi_+)}\prod_{\ell\neq j}^M\frac{(\lambda_j+\lambda_\ell)(\lambda_j-\lambda_\ell-1)}{(\lambda_j-\lambda_\ell+1)(\lambda_j+\lambda_\ell+2)}=1.
\label{BAE2}
\end{equation} 
Taking the derivative of the logarithm of the eigenvalue Eq.\eqref{TQrel}, we obtain
\begin{equation}
E=\frac{\partial\ln\Lambda(\lambda)}{\partial\lambda}\big\vert_{\lambda=0}-N=\sum_{j=1}^M \frac{2}{\lambda_j(\lambda_j+1)} +N-1+\frac{1}{\xi_-}+\frac{1}{\xi_+}.
\label{Engeqn}
\end{equation}
Changing the variables $\lambda_j=i\mu_j-\frac{1}{2}$, the energy Eq.\eqref{Engeqn} becomes
\begin{equation}
E=-\sum_{j=1}^M \frac{2}{\mu_j^2+\frac{1}{4}}+N-1+\frac{1}{\xi_-}+\frac{1}{\xi_+}.
\label{engeng}
\end{equation}
Similarly, the Bethe Ansatz equation Eq.\eqref{BAE2} can be written as
\begin{equation}
    \left(\frac{\mu_{j}-\frac{i}{2}}{\mu_{j}+\frac{i}{2}}\right)^{2 N} \left(\frac{\mu_{j}+i\left(\frac{1}{2}-\xi_-\right)}{\mu_{j}-i\left(\frac{1}{2}-\xi_-\right)}\right)\left(\frac{\mu_{j}+i\left(\frac{1}{2}-\xi_+\right)}{\mu_{j}-i\left(\frac{1}{2}-\xi_+\right)}\right) =\prod_{j \neq \ell=1}^{M}\left(\frac{\mu_{j}-\mu_{\ell}-i}{\mu_{j}-\mu_{\ell}+i}\right)\left(\frac{\mu_{j}+\mu_{\ell}-i}{\mu_{j}+\mu_{\ell}+i}\right).
\label{BAEfirst}
\end{equation}

From the Bethe equation Eq.\eqref{BAEfirst}, we can see that if $\mu_j\in\mathbb{C}$ is a solution, then $-\mu_j$ is also the solution. In other words, the distribution of roots of Bethe equations is symmetric about the origin.

\section{Analytic Solution of Bethe Ansatz Equation}
\label{sec:BAsolution}

Taking logarithm on both sides of Eq.\eqref{BAEfirst} and using $\ln\left(\frac{i-z}{i+z} \right)=2i\tan^{-1}(z)$, we get
\begin{align}
(2N+1)\tan^{-1}(2\mu_j)&-\tan^{-1}\left(\frac{\mu_j}{\frac{1}{2}-\xi_-}\right)-\tan^{-1}\left(\frac{\mu_j}{\frac{1}{2}-\xi_+}\right)\nonumber\\
&=\sum_{\ell=1}^M\left[\tan^{-1}(\mu_j-\mu_\ell)+\tan^{-1}(\mu_j+\mu_\ell) \right]+\pi I_j.
\label{FBAE}
\end{align}

The last term comes from the fact that $\exp$ function over $\mathbb{C}$ is not injective and hence its inverse function $\ln$ is multivalued. Here $I_j \in \mathbb{Z}$ acts as the quantum number of the model.

\begin{figure}[H]
  \centering
  \begin{subfigure}[b]{0.48\textwidth}
    \centering
    \includegraphics[width=7.5cm]{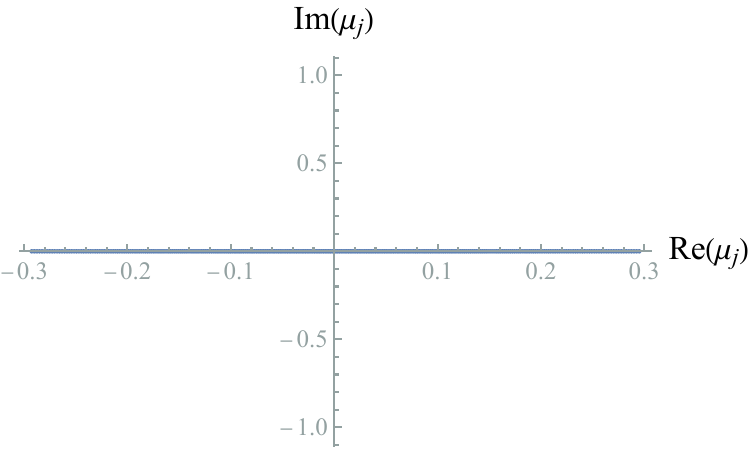}
    \subcaption{$\xi_-=-\frac{1}{4},\ \xi_+=-\frac{1}{3}$}
  \end{subfigure}\qquad
  \begin{subfigure}[b]{0.48\textwidth}
    \centering
    \includegraphics[width=7.5cm]{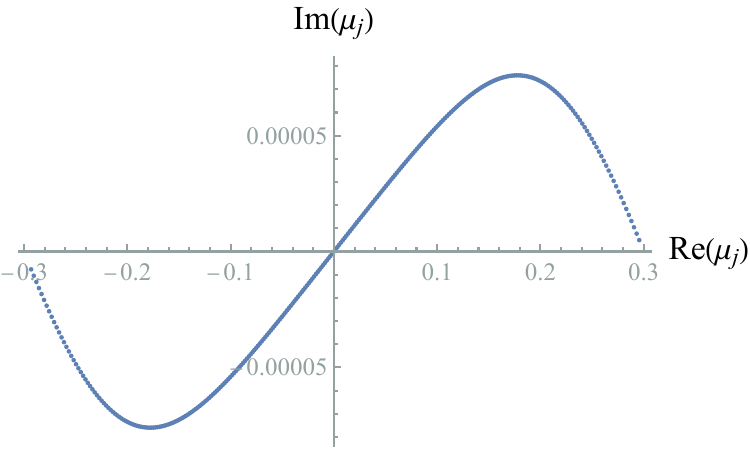}
    \subcaption{$\xi_-=\frac{1}{4}-\frac{i}{3},\ \xi_+=\frac{1}{3}-\frac{i}{5}$}
  \end{subfigure}

  \vspace{1em}

  \begin{subfigure}[b]{0.48\textwidth}
    \centering
    \includegraphics[width=7.5cm]{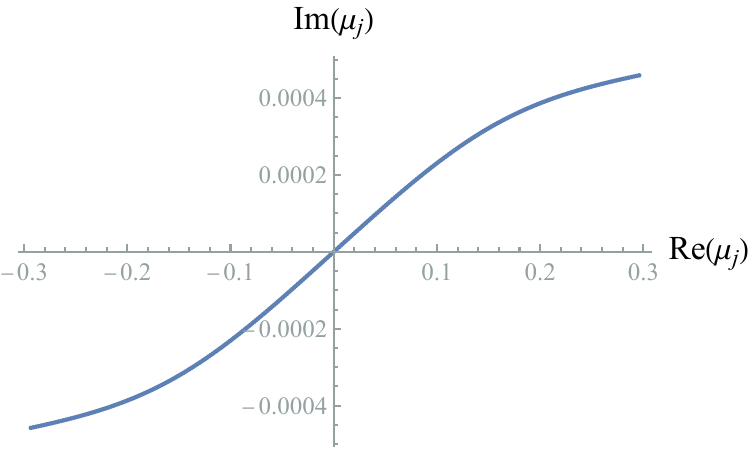}
    \subcaption{$\xi_-=\frac{1}{4}-\frac{i}{9},\ \xi_+=\frac{1}{3}-\frac{i}{7}$}
  \end{subfigure}\qquad
  \begin{subfigure}[b]{0.48\textwidth}
    \centering
    \includegraphics[width=7.5cm]{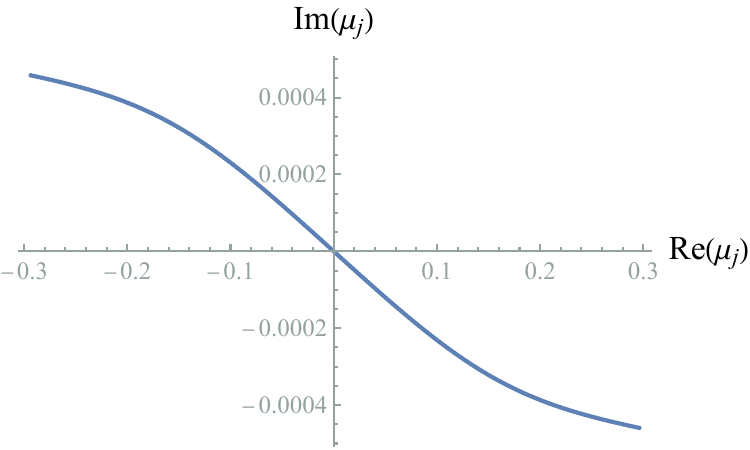}
    \subcaption{$\xi_-=\frac{1}{4}+\frac{i}{9},\ \xi_+=\frac{1}{3}+\frac{i}{7}$}
  \end{subfigure}

  \caption{Numerical solution of BAE \eqref{FBAE} for $M=N/2=300$ in the ground state.}
  \label{fig:numplots}
\end{figure}

To analyze Eq.\eqref{FBAE} in the thermodynamic limit, we define the density of Bethe roots as
\begin{equation}
\rho(\mu_j)=\frac{1}{\mu_{j+1}-\mu_j}.
\end{equation}
Such that we convert the sums over $j$ in Eq.\eqref{engeng} and Eq.\eqref{FBAE} into integral over $\mu$ as
\begin{equation}
E=-\int_\mathcal{C}\mathrm{d}\mu~ \rho(\mu)\frac{2}{\mu^2+\frac{1}{4}}+N-1+\frac{1}{\xi_-}+\frac{1}{\xi_+},\label{engcont}
\end{equation}
and
\begin{align}
(2N+1)\tan^{-1}(2\mu_j)&-\tan^{-1}\left(\frac{\mu_j}{\frac{1}{2}-\xi_-}\right)-\tan^{-1}\left(\frac{\mu_j}{\frac{1}{2}-\xi_+}\right)\nonumber\\
&=\int_\mathcal{C} \mathrm{d}\mu' \rho(\mu')\left[\tan^{-1}(\mu_j-\mu')+\tan^{-1}(\mu_j+\mu') \right]+\pi I_j,
\label{BAEcont}
\end{align}
where the integral is over the locus of $M-$coupled algebraic equations Eq.\eqref{BAEfirst}. If the parameters $h_{1/N}\in \mathbb{R}$ are real, then the locus would simply span the real line, and hence the limit of the integral would be $(-\infty,\infty)$. However when the parameters are allowed to be complex $h_{1/N}\in \mathbb{C}$, then the integration contour may lie in different regions of the complex plane depending on the boundary parameters as shown in Figure \ref{fig:numplots}.\\

We extract the density of roots in the ground state density $\rho_0(\mu)$ by subtracting Eq.\eqref{BAEcont} written for $\mu_j$ from the same equation written for $\mu_{j+1}$ and expanding in the difference $\Delta\mu=\mu_{j+1}-\mu_j$. This gives
\begin{equation}
2\rho_0(\mu)=f(\mu)-\int_\mathcal{C}\mathrm{d}\mu~ K(\mu-\mu')\rho_0(\mu')-\int_\mathcal{C}\mathrm{d}\mu~ K(\mu+\mu')\rho_0(\mu')+\mathcal{O}\left(\frac{1}{N}\right).
\label{SolDensity}
\end{equation}
Since we are interested in thermodynamics limit $N\to \infty$, the higher order terms are negligible. Here,
\begin{align}
f(\mu)&=(2N+1)a_{\frac12}(\mu)-a_{\frac{1}{2}-\xi_-}(\mu)-a_{\frac{1}{2}-\xi_+}(\mu), \\
a_\gamma(\mu)&=\frac{1}{\pi}\frac{\gamma}{\mu^2+\gamma^2}, \quad\text{and}\\
K(\mu)&=\frac{1}{\pi  \left(\mu ^2+1\right)}=a_1(\mu)\label{kernel}
\end{align}

We need to solve the integral equation Eq.\eqref{SolDensity} in order to find the eigenvalues given by Eq.\eqref{engcont}. However, it is very difficult to solve the coupled algebraic Bethe Ansatz Eq.\eqref{BAEfirst} exactly to find its locus, over which we need to perform the integration to find the root density. In order to get around this problem, we choose $\mathscr{PT}$-symmetric boundary fields $h_1=\frac{1}{\xi_-}=\frac{1}{\xi+i\chi}$ and $h_N=\frac{1}{\xi_+}=\frac{1}{\xi-i\chi}$ where $\{\xi,\chi\}\in\mathbb{R}$. Let us recall the action of time reversal and parity in a discrete system; the time reversal operation $\mathscr{T}$ is such that $\mathscr{T} i \mathscr{T}=-i$ and the effect of the parity on a system of $N$ spins is such that $\mathscr{P} \sigma_{l}^{\alpha} \mathscr{P}=\sigma_{N+1-l}^{\alpha}$. Thus, a $\mathscr{PT}$-symmetric $XXX$ model has the Hamiltonian of the form
\begin{equation}
\mathcal{H}=\sum_{j=1}^{N-1}\sum_{\alpha=1}^3\sigma_j^\alpha\sigma_{j+1}^\alpha+\frac{1}{\xi+i\chi} \sigma_1^z+\frac{1}{\xi-i\chi}\sigma_N^z.
\end{equation}

One can immediately see that the two $\tan^{-1}$ functions in the first line of Eq.\eqref{FBAE} combine to have a real argument and hence the solution of the logarithmic Bethe Ansatz is real just as in the real case. One can numerically look at the solution of Eq.\eqref{FBAE} with the $\mathscr{PT}$-symmetric boundary fields.
\begin{figure}[H]
    \centering
    \includegraphics{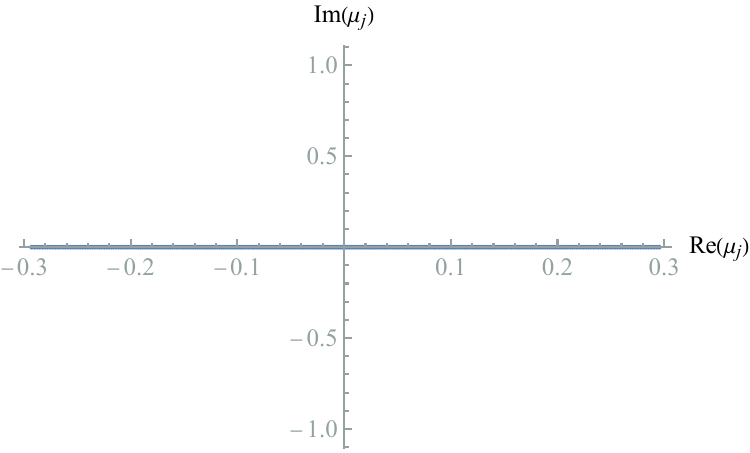}
    \caption{Distribution of roots in ground state for the choice $\xi_\pm=\frac{1}{4}\pm \frac{i}{3}$.}
    \label{fig:pt}
\end{figure}

Since the roots lie in the real line as shown in Figure \ref{fig:pt}, we can take the integration over the entire real line in Eq.\eqref{SolDensity} and solve for the density distribution of Bethe roots.

\subsection{Boundary Strings Solution}
It is well known that the solutions of the Bethe Ansatz equations for arbitrary $M$, in the thermodynamics limit $N\to\infty$, contain spin complexes of various lengths $\ell$. An $\ell-$spin complex
consists of an aggregate of $\ell$ complex rapidities $\mu_j$, which all have a common real part and whose imaginary parts differ by unity. The spin complex is thus described by $\ell$ complex numbers,
\begin{equation}
    \mu^{\ell, j}=\mu^\ell+\frac{i}{2}(\ell+1-2 j), \quad j=1,2, \ldots, \ell
\end{equation}
where $\mu^\ell \in \mathbb{R}$. These solutions describe the complex bound states of magnetic excitations. \\

From the Bethe Ansatz equation Eq.\eqref{BAEfirst}, it is easy to see that when $\xi<\frac{1}{2}$, there are two additional solutions in the thermodynamic limit. We call these boundary string solutions. The left boundary string is of the form
\begin{equation}
\mu_{L}=\pm i\left(\frac{1}{2}+(\xi+i\chi)\right),
\end{equation}
and the right boundary string is of the form
\begin{equation}
\mu_{R}=\pm i\left(\frac{1}{2}+(\xi-i\chi)\right).
\end{equation}
  The parametric region $|\xi|<\frac{1}{2}$ where boundary string solutions exist are henceforth called A Phases and the region $|\xi|>\frac{1}{2}$ where boundary solution do not exist are called B phases. We solve the Bethe Ansatz Equation in different boundary parametric regimes for odd and even numbers of sites separately and list all the elementary excitations. 
\subsection{A Phases}
\subsubsection{{\color{blue}{$A_1 $ Phase}}}:
Consider the phase $A_1$ where the real part of the boundary parameter takes the value $0<\xi<\frac{1}{2}$. The Bethe equation corresponding to the reference state with all spins down is obtained from Eq.\eqref{BAEfirst} after applying the transformation Eq.\eqref{symeqn} \textit{i.e.}
\begin{equation}
\left(\frac{\mu_{j}-\frac{i}{2}}{\mu_{j}+\frac{i}{2}}\right)^{2 N} \left(\frac{\mu_{j}-\chi+i \left(\frac 1 2+\xi\right)}{\mu_{j}+\chi-i\left(\frac{1}{2 }+\xi\right)}\right)\left(\frac{\mu_{j}+\chi+i\left(\frac{1}{2 }+\xi\right)}{\mu_{j}-\chi-i\left(\frac{1}{2 }+\xi\right)}\right) =\prod_{j \neq \ell=1}^{M}\left(\frac{\mu_{j}-\mu_{\ell}-i}{\mu_{j}-\mu_{\ell}+i}\right)\left(\frac{\mu_{j}+\mu_{\ell}-i}{\mu_{j}+\mu_{\ell}+i}\right).
\label{bbeqbr}
\end{equation}
The eigenvalue of the Hamiltonian is given by
\begin{equation}
E=-\sum_{j=1}^M \frac{2}{\mu_j^2+\frac{1}{4}}+N-1-\frac{1}{\xi+i\chi}-\frac{1}{\xi-i\chi}=-\sum_{j=1}^M \frac{2}{\mu_j^2+\frac{1}{4}}+N-1-\frac{2 \xi }{\xi ^2+\chi ^2}.
\end{equation}
Due to the $\mathscr{PT}$-symmetry that the energy is real even though the Hamiltonian is not Hermitian.

\underline{\centering{$A_1$ Phase: Odd number of sites}}\\

The ground state root density is obtained from Eq.\eqref{SolDensity}, which is given by
\begin{equation}
2\rho_{\left|-\frac12\right\rangle}(\mu)=(2N+1)a_{\frac12}(\mu)-a_{\frac{1}{2}+\xi+i\chi}(\mu)-a_{\frac{1}{2}+\xi-i\chi}(\mu)-\sum_{\upsilon=\pm}\int_{-\infty}^\infty\mathrm{d}\mu~ a_1(\mu+\upsilon\mu')\rho_{\left|-\frac12\right\rangle}(\mu')-\delta(\mu).
\end{equation}
Where the delta function is added to remove the solution $\mu_j=0$ solution which results in a vanishing wavefunction \cite{grisaru1995direct}.\\
 The solution of the above equation in the Fourier space takes the following form
\begin{equation}
\tilde\rho_{\left|-\frac12\right\rangle}(\omega)=\frac{(2 N+1) e^{-\frac{| \omega | }{2}}-e^{- \left(\frac12+ \xi +i\chi\right) | \omega | }-e^{- \left( \frac12+\xi-i\chi\right) | \omega | }-1}{2 \left(e^{-| \omega | }+1\right)}
.\end{equation}
The total number of Bethe roots is given by
\begin{equation}
M_{\left|-\frac12\right\rangle}=\tilde\rho_{\left|-\frac12\right\rangle}(\omega=0)=\frac{N-1}{2}.
\end{equation}
We can compute the $z-$component of the total spin, where
\begin{equation}
S^z_{\left|-\frac12\right\rangle}=-\left(\frac{N}{2}-M_{\left|-\frac12\right\rangle}\right)=-\frac{1}{2}.
\end{equation}

To compute the energy using Eq.\eqref{engcont}, we need to compute the integral
\begin{align}
-\int_{-\infty}^\infty \mathrm{d}\mu \frac{2\rho_{\left|-\frac12\right\rangle}(\mu)}{\mu^2+\frac{1}{4}}=\pi&-(2 N+1) \ln (4)+\psi ^{(0)}\left(\frac{\xi+i\chi}{2}+1\right)-\psi ^{(0)}\left(\frac{\xi+i\chi}{2}+\frac{1}{2}\right)\nonumber\\
&+\psi ^{(0)}\left(\frac{\xi-i\chi}{2}+1\right)-\psi ^{(0)}\left(\frac{\xi-i\chi}{2}+\frac{1}{2}\right),
\end{align}
where $\psi^0(z)=\frac{d}{dz}\left(\ln\Gamma(z)\right)$ is the diGamma function. 

Thus, the energy of this state is
\begin{align}
E_{\left|-\frac12\right\rangle}=E_0=&N-1+\pi-(2 N+1) \ln (4)-\frac{1}{\xi+i\chi}-\frac{1}{\xi-i\chi}+\psi ^{(0)}\left(\frac{\xi+i\chi}{2}+1\right)\nonumber\\
&-\psi ^{(0)}\left(\frac{\xi+i\chi}{2}+\frac{1}{2}\right)+\psi ^{(0)}\left(\frac{\xi-i\chi}{2}+1\right)-\psi ^{(0)}\left(\frac{\xi -i\chi}{2}+\frac{1}{2}\right)\in\mathbb{R}.
\label{engeqnreim}
\end{align}

Adding the boundary string solution $\mu_L=\pm i\left(\frac{1}{2}+(\xi+i\chi)\right)$, we get
\begin{align}
2\rho_{\ket{0}_L}(\mu)&=(2N+1)a_{\frac12}(\mu)-a_{\frac{1}{2}+\xi+i\chi}(\mu)-a_{\frac{1}{2}+\xi-i\chi}(\mu)-a_{\frac{1}{2}-\xi-i\chi}(\mu)-a_{\frac{3}{2}+\xi +i \chi }(\mu)\nonumber\\
&-2\int_{-\infty}^\infty\mathrm{d}\mu~ a_1(\mu-\mu')\rho_{\ket{0}_L}(\mu')-\delta(\mu).
\end{align}
In the Fourier space, we obtain the solution
\begin{equation}
\tilde\rho_{{\left|0\right\rangle}_L}(\omega)=\tilde{\rho}_{\left|-\frac12\right\rangle}(\omega)+\Delta\tilde\rho_{L}(\omega) \quad \text{where}~ \Delta\tilde\rho_{L}(\omega)=-\frac{e^{-\left(\frac{1}{2}-\xi -i \chi \right)| \omega | }+e^{-\left(\frac{3}{2}+\xi +i \chi \right)| \omega | }}{2 \left(e^{-| \omega | }+1\right)}.
\end{equation}
The total number of roots is given by
\begin{equation}
M_L=1+\tilde\rho_{0_L}(0)=1+\frac{N-1}{2}-\frac{1}{2}=\frac{N}{2}.
\end{equation}
Hence, the $z-$component of the total spin is 
\begin{equation}
S^z_{L}=0.
\end{equation}
Notice that the number of roots is an integer only if the number of sites is even. Since we have a chain with an odd number of sites, in order for one to add a boundary string to the above state, a propagating hole (spinon) needs to be added. Adding a spinon with rapidity $\theta$, we obtain
\begin{align}
2\rho_{{\left|-\frac12\right\rangle}_{\theta,L}}(\mu)=(2N+1)& a_{\frac12}(\mu)- a_{\frac{1}{2}+\xi+i\chi}(\mu)-a_{\frac{1}{2}+\xi-i\chi}(\mu)-a_{\frac{1}{2}-\xi-i\chi}(\mu)-a_{\frac{3}{2}+\xi +i \chi }(\mu)\nonumber\\
&-2\int_{-\infty}^\infty\mathrm{d}\mu~ a_1(\mu-\mu')\rho_{{\left|-\frac12\right\rangle}_{\theta,L}}(\mu')-\delta(\mu)-\delta(\mu-\theta)-\delta(\mu+\theta).
\end{align}
The solution of the above equation in the Fourier space is immediate
\begin{equation}
\tilde\rho_{{\left|-\frac12\right\rangle}_{\theta,L}}(\omega)=\tilde{\rho}_{\left|-\frac12\right\rangle}(\omega)+\Delta\tilde\rho_{L}(\omega)+\Delta\tilde\rho_\theta(\omega) \quad \text{where}~ \Delta\tilde\rho_{\theta}(\omega)=\frac{-e^{-i \theta  \omega }-e^{i \theta  \omega }}{2 \left(e^{-| \omega | }+1\right)}=-\frac{\cos(\theta\omega)}{\left(e^{-| \omega | }+1\right)}.
\end{equation}
The total number of real roots is given by
\begin{equation}
M_{{\left|-\frac12\right\rangle}_{\theta,L}}=1+\tilde{\rho}_{{\left|-\frac12\right\rangle}_{\theta,L}}(0)=\frac{N-1}{2}.
\end{equation}
Thus, the $z-$component of the total spin is
\begin{equation}
S_{\theta,L}^z=-\left(\frac{N}{2}-\left(\frac{N}{2}-\frac{1}{2}\right)\right)=-\frac12.
\end{equation}
We can compute the energy of the system using Eq.\eqref{engcont}, and we obtain
\begin{equation}
E_{{\left|-\frac12\right\rangle}_{\theta,L}}=-\int_{-\infty}^\infty\mathrm{d}\mu~ \rho_{{\left|-\frac12\right\rangle}_{\theta,L}}(\mu)\frac{2}{\mu^2+\frac{1}{4}}+N-1+\frac{1}{\xi+i\chi}+\frac{1}{\xi-i\chi}-\frac{2}{\frac{1}{4}+\left(i\left(\frac{1}{2}+(\xi+i\chi)\right)\right)^2}.
\label{tteng}
\end{equation}
Using  Eq.\eqref{engeqnreim}, the energy given by Eq.\eqref{tteng} can be simplified as
\begin{equation}
E_{{\left|-\frac12\right\rangle}_{\theta,L}}=E_0+\frac{2\pi}{\sin(\pi(\xi+i\chi))}+\frac{2\pi}{\cosh(\pi\theta)}
\label{englts}.
\end{equation}
The first term is the energy of the ground state $\left|-\frac12\right\rangle$, the second term 
\begin{equation}
    m=\frac{2\pi}{\sin(\pi(\xi+i\chi))}
\end{equation}
is the energy of the bound state at the left boundary and the third term
\begin{equation}
    E_\theta=\frac{2\pi}{\cosh(\pi\theta)}
\label{energyspinon}
\end{equation}
is the energy of the spinon propagating with rapidity $\theta$. Notice that the first and the third term are real but the second term is complex.

The real part of the energy is
\begin{equation}
\Re \left(E_{{\left|-\frac12\right\rangle}_{\theta,L}}\right)=E_0+\frac{2\pi}{\cos(\pi\theta)}+\frac{4 \pi  \sin (\pi  \xi ) \cosh (\pi  \chi )}{\cosh (2 \pi  \chi )-\cos (2 \pi  \xi )}
\label{boundenrealleft}
\end{equation}
The energy of the spinon is strictly positive with energy range $0<E_\theta<2\pi$. Likewise, the real part of the energy of the bound state is also strictly positive  with a range $E_L>0$. { When $\chi=0$, the lowest value of the real part of the energy of the bound state is $E_{L}^-=2\pi$, but as $\chi\to\infty$, the lowest value of the real part of the energy vanishes.}

The imaginary part of the energy is 
\begin{equation}
\Im \left(E_{{\left|-\frac12\right\rangle}_{\theta,L}}\right)=-\frac{4 \pi  \cos (\pi  \xi ) \sinh (\pi  \chi )}{\cosh (2 \pi  \chi )-\cos (2 \pi  \xi )}.
\label{boundenimagleft}
\end{equation}
{{Since $0<\xi<\frac{1}{2}$ and $\chi>0$,  the imaginary part of the equation is always negative. Exactly at $\xi=\frac12$, the imaginary part of the energy vanishes for all values of $\chi$, and at $\chi=0$ and $\chi=\infty$, the imaginary part of the energy vanishes for all values of $\xi$. The negativity of the imaginary part of the energy suggests that there is loss at the left boundary of the system.}}

We can obtain the state $\left|-\frac{1}{2}\right\rangle_{\theta,R}$ by adding the right boundary string and a spinon with rapidity $\theta$. This state is described by following the density distribution
\begin{equation}
\tilde\rho_{{\left|-\frac12\right\rangle}_{\theta,R}}(\omega)=\tilde{\rho}_{\left|-\frac12\right\rangle}(\omega)+\Delta\tilde\rho_{R}(\omega)+\Delta\tilde\rho_\theta(\omega) \quad \text{where}~ \; \Delta\tilde\rho_{R}(\omega)=-\frac{e^{-\left(\frac{1}{2}-\xi +i \chi \right)| \omega | }+e^{-\left(\frac{3}{2}+\xi -i \chi \right)| \omega | }}{2 \left(e^{-| \omega | }+1\right)},
\end{equation}
and its energy is given by
\begin{equation}
E_{{\left|-\frac12\right\rangle}_{\theta,R}}=E_0+\frac{2\pi}{\sin(\pi(\xi-i\chi))}+\frac{2\pi}{\cosh(\pi\theta)}.
\end{equation}
Notice that this is exactly the complex conjugate of the energy Eq.\eqref{englts} of the state $\ket{-\frac{1}{2}}_{\theta,L}$ constructed above. Once again, the first term is the energy of the ground state configuration, the third term is the energy of the spinon, and the second term
\begin{equation}
    m^*=\frac{2\pi}{\sin(\pi(\xi-i\chi))}
\end{equation}
is the energy of the boundary mode situated at the right end of the chain.\\
The real part of the energy is
\begin{equation}
\Re \left(E_{{\left|-\frac12\right\rangle}_{\theta,R}}\right)=E_0+\frac{2\pi}{\cos(\pi\theta)}+\frac{4 \pi  \sin (\pi  \xi ) \cosh (\pi  \chi )}{\cosh (2 \pi  \chi )-\cos (2 \pi  \xi )},
\label{boundenrealright}
\end{equation}
and the imaginary part of the energy is
\begin{equation}
\Im \left(E_{{\left|-\frac12\right\rangle}_{\theta,R}}\right)=\frac{4 \pi  \cos (\pi  \xi ) \sinh (\pi  \chi )}{\cosh (2 \pi  \chi )-\cos (2 \pi  \xi )}.
\label{boundenimagright}
\end{equation}
Notice that the imaginary part of the energy of the bound state at the right boundary is exactly equal to the negative of the imaginary part of the energy of the bound state at the left boundary. The imaginary part of the energy is strictly positive and hence there is gain at the right boundary of the system.

Now, we consider the Bethe Ansatz equation corresponding to the reference state with all spins up given by Eq.\eqref{BAEfirst}.

The eigenvalues of the Hamiltonian are given by
\begin{equation}
E=-\sum_{j=1}^M \frac{2}{\mu_j^2+\frac{1}{4}}+N-1+\frac{1}{\xi+i\chi}+\frac{1}{\xi-i\chi}=-\sum_{j=1}^M \frac{2}{\mu_j^2+\frac{1}{4}}+N-1+\frac{2 \xi }{\xi ^2+\chi ^2}.
\end{equation}
Taking the logarithm of the Bethe Ansatz equation, we extract the density distribution of the Bethe roots using the same process as above. We obtain the following density distribution in the Fourier space 
\begin{equation}
\tilde\rho_{\left|\frac12\right\rangle}(\omega)=\frac{(2 N+1) e^{-\frac{| \omega | }{2}}-e^{- \left(\frac12- \xi -i\chi\right) | \omega | }-e^{- \left( \frac12-\xi+i\chi\right) | \omega | }-1}{2 \left(e^{-| \omega | }+1\right)}.
\label{lrdist}
\end{equation}
The total number of Bethe roots is given by
\begin{equation}
M_{\left|\frac12\right\rangle}=\tilde\rho_{\left|\frac12\right\rangle}(\omega=0)=\frac{N-1}{2}.
\end{equation}
We can compute the $z-$component of the total spin, where
\begin{equation}
S^z=\left(\frac{N}{2}-M_{\left|-\frac12\right\rangle}\right)=\frac{1}{2}.
\end{equation}
The energy of the state is given by
\begin{equation}
E_{\left|\frac{1}{2}\right\rangle}=-\int_{-\infty}^\infty\mathrm{d}\mu~ \rho_{\left|\frac{1}{2}\right\rangle}(\mu)\frac{2}{\mu^2+\frac{1}{4}}+N-1+\frac{1}{\xi+i\chi}+\frac{1}{\xi-i\chi}.
\label{engpoh}
\end{equation}
Writing
\begin{equation}
\tilde\rho_{\left|\frac12\right\rangle}(\omega)=\tilde\rho_{-\left|\frac12\right\rangle}(\omega)+\tilde\rho_*(\omega),
\label{defrh}
\end{equation}
where
\begin{equation}
\tilde\rho_*(\omega)=\frac{e^{ \left(\frac12+ \xi+i\chi\right) | \omega | }-e^{- \left(\frac12- \xi -i\chi\right) | \omega | }+e^{- \left( \frac12+\xi-i\chi\right) | \omega | }-e^{- \left( \frac12-\xi+i\chi\right) | \omega | }}{2 \left(e^{-| \omega | }+1\right)},
\label{wrdtrk}
\end{equation}
and using Eq.\eqref{defrh} and Eq.\eqref{wrdtrk}, we can write the energy of the state $\left|\frac12\right\rangle$ given by Eq.\eqref{engpoh} as
\begin{equation}
E_{\left|\frac{1}{2}\right\rangle}=E_{\left|-\frac{1}{2}\right\rangle}+\frac{2}{\xi+i\chi}+\frac{2}{\xi-i\chi}-\int_{-\infty}^\infty\mathrm{d}\mu~ \rho_*(\mu)\frac{2}{\mu^2+\frac{1}{4}}.
\end{equation}
By computing the integral
\begin{equation}
-\int_{-\infty}^\infty\mathrm{d}\mu~ \rho_*(\mu)\frac{2}{\mu^2+\frac{1}{4}}=2 \pi  \csc (\pi  (\xi +i \chi ))-\frac{2}{\xi +i \chi }+2 \pi  \csc (\pi  (\xi -i \chi ))-\frac{2}{\xi -i \chi },
\end{equation}
 we obtain
\begin{equation}
E_{\left|\frac{1}{2}\right\rangle}=E_{\left|-\frac{1}{2}\right\rangle}+\frac{2\pi}{\sin(\pi(\xi+i\chi))}+\frac{2\pi}{\sin(\pi(\xi-i\chi))}.
\label{totlreng}
\end{equation}
The last two terms are the energies of the bound state at the left boundary and right boundary respectively. Thus, this state contains a bound state at both the left and the right boundaries. We will henceforth represent this state as $\left|\frac12\right\rangle_{L,R}$. We can rewrite the energy as
\begin{equation}
E_{{\left|\frac{1}{2}\right\rangle}_{L,R}}=E_{\left|-\frac{1}{2}\right\rangle}+\frac{8 \pi  \sin (\pi  \xi ) \cosh (\pi  \chi )}{\cosh (2 \pi  \chi )-\cos (2 \pi  \xi )},
\end{equation}
which shows that the energy of the state is real and strictly positive.\\

For the BAE with reference state with all spins up, the boundary strings are of the form
\begin{equation}
\mu_{L'}=\pm i\left(\frac{1}{2}-(\xi+i\chi)\right)
\end{equation}
and
\begin{equation}
\mu_{R'}=\pm i\left(\frac{1}{2}-(\xi-i\chi)\right).
\end{equation}
Now, we can construct a state ${\left|\frac{1}{2}\right\rangle}_{\theta,L}$ by adding a spinon and the boundary string $\mu_{R'}=\pm i\left(\frac{1}{2}-(\xi-i\chi)\right)$ to Eq.\eqref{lrdist} to obtain
\begin{equation}
\tilde{\rho}_{{\left|\frac{1}{2}\right\rangle}_{\theta,L}}(\omega)=\tilde{\rho}_{\left|\frac{1}{2}\right\rangle}(\omega)+\Delta\tilde\rho_\theta(\omega)+\Delta\tilde\rho_{R'}(\omega) \quad\text{where}~ \Delta\tilde\rho_{R'}(\omega)=-\frac{e^{-\left(\frac{1}{2}+\xi -i \chi \right)| \omega | }+e^{-\left(\frac{3}{2}-\xi +i \chi \right)| \omega | }}{2 \left(e^{-| \omega | }+1\right)}.
\end{equation}
The total number of roots is given by 
\begin{equation}
M_{{\left|\frac12\right\rangle}_{\theta,L}}=1+\tilde{\rho}_{{\left|\frac12\right\rangle}_{\theta,L}}(0)=\frac{N-1}{2}.
\end{equation}
Thus, the $z-$component of the total spin is
\begin{equation}
S_{\theta,L}^z=\left(\frac{N}{2}-\left(\frac{N}{2}-\frac{1}{2}\right)\right)=\frac12.
\end{equation}
The energy of this state is given by
\begin{equation}
E_{{\left|\frac{1}{2}\right\rangle}_{\theta,L}}=E_{\left|\frac12\right\rangle}+\frac{2\pi}{\cosh(\pi\theta)}-\frac{2\pi}{\sin(\pi(\xi+i\chi))}.
\end{equation}
Using Eq.\eqref{totlreng}, we get
\begin{equation}
E_{{\left|\frac{1}{2}\right\rangle}_{\theta,L}}=E_{\left|-\frac12\right\rangle}+\frac{2\pi}{\cosh(\pi\theta)}+\frac{2\pi}{\sin(\pi(\xi-i\chi))}
\end{equation}
This state contains a bound state at the left boundary and is degenerate with the state $\left|-\frac12\right\rangle_{\theta,L}$.

If we add a spinon with rapidity $\theta$ and the boundary string $\mu_{L'}=\pm i\left(\frac{1}{2}-(\xi+i\chi)\right)$ to the state $\left|\frac12\right\rangle$, we obtain the state ${\left|\frac{1}{2}\right\rangle}_{\theta,R}$ which contains a bound state at the right boundary and is degenerate to the state ${\left|-\frac{1}{2}\right\rangle}_{\theta,R}$. All other excited states can be constructed by adding bulk excitations with purely real energy,  such as an even number of spinons, bulk strings, and quartets. Thus, adding these bulk excitations on top of a $\mathscr{PT}$ symmetric elementary excitations with real energies like $\left|-\frac{1}{2}\right\rangle$ and $\left|\frac{1}{2}\right\rangle_{L/R}$, we construct excited states with real energies. Whereas, adding the bulk excitations on top of $\mathscr{PT}$ broken states $\left|\pm\frac{1}{2}\right\rangle_{\theta,R}$ and $\left|\pm\frac{1}{2}\right\rangle_{\theta,L}$, we create excited states in $\mathscr{PT}$ broken phase. In the thermodynamic limit, for each excited state built on top of state with bound state on the right edge, there exists a corresponding excited state built on top of state with bound state on the left edge whose energies are complex conjugates of each other.  

\underline{\centering{$A_1$ Phase: Even number of sites}}\\
Recall that the ground state configuration of the spin chain with an odd number of sites has the root density 
\begin{equation}
{\tilde\rho_{\left|-\frac12\right\rangle}}(\omega)=\frac{(2 N+1) e^{-\frac{| \omega | }{2}}-e^{- \left(\frac12+ \xi +i\chi\right) | \omega | }-e^{- \left( \frac12+\xi-i\chi\right) | \omega | }-1}{2 \left(e^{-| \omega | }+1\right)},
\label{gsdist}
\end{equation}
thus, the number of roots is $M=\frac{N-1}{2}$. If $N$ is even, the number of roots is not an integer. Thus, to consider the ground state configuration of the spin chain with an even number of sites, we need to add a spinon to Eq. \eqref{gsdist} \textit{i.e.} consider the following distribution
\begin{equation}
\tilde\rho_{{\left|-1\right\rangle}_\theta}(\omega)=\frac{(2 N+1) e^{-\frac{| \omega | }{2}}-e^{- \left(\frac12+ \xi +i\chi\right) | \omega | }-e^{- \left( \frac12+\xi-i\chi\right) | \omega | }-1}{2 \left(e^{-| \omega | }+1\right)}+\Delta\tilde\rho_\theta(\omega), 
\label{gsevndist}
\end{equation}
where
$ \Delta\tilde\rho_\theta(\omega)=-\frac{\cos(\omega\theta)}{(1+e^{-|\omega|}}$. The total number of roots is given by
\begin{equation}
M_{{\left|-1\right\rangle}_\theta}=\tilde\rho_{{\left|-1\right\rangle}_\theta}(\omega=0)=\frac{N-2}{2}.
\end{equation}
Clearly, the number of roots is an integer only for the spin chain with an even number of sites. The $z-$component of the total spin of this state is
\begin{equation}
S^z=-\left(\frac{N}{2}-M_{{\left|-1\right\rangle}_\theta}\right)=-1.
\end{equation}
Using Eq.\eqref{engcont}, we can compute the energy of this state and we obtain
\begin{equation}
E_{{\left|-1\right\rangle}_\theta}= E_0+\frac{2\pi}{\cosh(\pi\theta)}.
\end{equation}
The lowest energy state for a spin chain with an even number of sites is parameterized by the rapidity $\theta$ of the spinon. Since $\cosh$ is a monotonically non-decreasing function, the ground state is obtained in the limit $\theta\to \infty$.

Next, considering the reference state with all spins down, we take the state with all real roots and add the left boundary string solution $\mu_L=\pm i\left(\frac{1}{2}+(\xi+i\chi)\right)$, which gives the root density of the form
\begin{align}
\tilde\rho_{{\left|0\right\rangle}_L}(\omega)&=\frac{(2 N+1) e^{-\frac{| \omega | }{2}}-e^{- \left(\frac12+ \xi +i\chi\right) | \omega | }-e^{- \left( \frac12+\xi-i\chi\right) | \omega | }-1}{2 \left(e^{-| \omega | }+1\right)}+\Delta\tilde\rho_{L}(\omega),\\
& \quad \text{where }~ \Delta\tilde\rho_{L}(\omega)=-\frac{e^{-\left(\frac{1}{2}-\xi -i \chi \right)| \omega | }+e^{-\left(\frac{3}{2}+\xi +i \chi \right)| \omega | }}{2 \left(e^{-| \omega | }+1\right)}\nonumber.
\end{align}

The total number of roots is given by
\begin{equation}
M_{\ket{0}_L}=1+\tilde\rho_{0_L}(0)=1+\frac{N-1}{2}-\frac{1}{2}=\frac{N}{2}.
\end{equation}
Hence, the $z-$component of the total spin is 
\begin{equation}
S^z_{\ket{0}_L}=0.
\end{equation}
The energy of this state henceforth represented at $\ket{0}_L$ is 
\begin{equation}
E_{\ket{0}_L}=E_0+\frac{2\pi}{\sin(\pi(\xi+i\chi))}.
\label{0Leng}
\end{equation}

Likewise, we can obtain the state $\ket{0}_R$ with energy
\begin{equation}
E_{\ket{0}_R}=E_0+\frac{2\pi}{\sin(\pi(\xi-i\chi))},
\label{0Reng}
\end{equation}
by adding right boundary string solution $\mu_R=\pm i\left(\frac{1}{2}+(\xi-i\chi)\right)$ to the state containing all real roots with respect to the reference state with all spins down.

Let us add both string solutions and a spinon to create a state with root distribution
\begin{equation}
\tilde{\rho}_{\ket{0}_{\theta,L,R}}(\omega)=\frac{(2 N+1) e^{-\frac{| \omega | }{2}}-e^{- \left(\frac12+ \xi +i\chi\right) | \omega | }-e^{- \left( \frac12+\xi-i\chi\right) | \omega | }-1}{2 \left(e^{-| \omega | }+1\right)}+\Delta\tilde\rho_{L}(\omega)+\Delta\tilde\rho_{R}(\omega)+\Delta\tilde\rho_{\theta}(\omega).
\end{equation}
The total number of roots is given by
\begin{equation}
M_{\ket{0}_{\theta,L,R}}=2+\rho_{\ket{0}_{\theta,L,R}}=\frac{N}{2},
\end{equation}
And hence the $z-$component of the spin is 
\begin{equation}
S^z_{\ket{0}_{\theta,L,R}}=0.
\end{equation}
The energy of the state is given by
\begin{equation}
E_{\ket{0}_{\theta,L,R}}=E_0+\frac{2\pi}{\sin(\pi(x+i\chi))}+\frac{2\pi}{\sin(\pi(\xi-i\chi))}+\frac{2\pi}{\cosh(\pi\theta)}
\label{tlr0}.
\end{equation}

Now, consider the state with all real roots corresponding to the reference state with all spins up and a spinon described by the density
\begin{equation}
\tilde\rho_{{\left|1\right\rangle}_\theta}(\omega)=\frac{(2 N+1) e^{-\frac{| \omega | }{2}}-e^{- \left(\frac12- \xi -i\chi\right) | \omega | }-e^{- \left( \frac12-\xi+i\chi\right) | \omega | }-1}{2 \left(e^{-| \omega | }+1\right)}+\Delta\tilde\rho_\theta(\omega).
\end{equation}
The total number of roots is given by
\begin{equation}
M_{{\left|1\right\rangle}_\theta}=\tilde\rho_{{\left|1\right\rangle}_\theta}(\omega=0)=\frac{N-2}{2},
\end{equation}
which is an integer only for the even number of sites. And the $z-$component of the spin of this state is
\begin{equation}
S^z=\left(\frac{N}{2}-M_{{\left|1\right\rangle}_\theta}\right)=1.
\end{equation}
We can compute the energy of this state, and we obtain
\begin{equation}
E_{\ket{1}_\theta}=E_0+\frac{2\pi}{\sin(\pi(x+i\chi))}+\frac{2\pi}{\sin(\pi(\xi-i\chi))}+\frac{2\pi}{\cosh(\pi\theta)}.
\end{equation}
This state contains a spinon and bound states at both boundaries. This state represented as $\ket{1}_{\theta,L,R}$ henceforth is degenerate with the state $\ket{0}_{\theta,L,R}$. 

We can add both boundary strings $\mu_{L'}=\pm i\left(\frac{1}{2}-(\xi+i\chi)\right)$ and $\mu_{R'}=\mu_{L'}=\pm i\left(\frac{1}{2}-(\xi-i\chi)\right)$ and a spinon to the state with all real roots with respect to the reference state with all spins up. We obtain a state with the following root density
\begin{align}
\tilde{\rho}_{\ket{0}_\theta}(\omega)=&\frac{(2 N+1) e^{-\frac{| \omega | }{2}}-e^{- \left(\frac12- \xi -i\chi\right) | \omega | }-e^{- \left( \frac12-\xi+i\chi\right) | \omega | }-1}{2 \left(e^{-| \omega | }+1\right)}\nonumber\\
&\quad+\Delta\tilde\rho_\theta(\omega)+\Delta\tilde\rho_\theta(\omega)+\Delta\tilde{\rho}_{R'}(\omega)+\Delta\tilde\rho_{L'}(\omega).
\end{align}

The total number of roots in this state is
\begin{equation}
M_{\ket{0}_\theta}=2+\tilde{\rho}_{\ket{0}_\theta}(0)=\frac{N}{2},
\end{equation}
and the $z-$component of the spin of this state is equal to
\begin{equation}
S_{\ket{0}_\theta}=0.
\end{equation}
The energy of this state is given by
\begin{equation}
E_{\ket{0}_\theta}=E_0+\frac{2\pi}{\cosh(\pi\theta)},
\end{equation}
and is degenerate with the state $\ket{-1}_\theta$.

Note that the state $\ket{0}_L$ with energy given by Eq.\eqref{0Leng} can also be created by starting with all the real roots with respect to the reference state with all spins up and adding the right boundary string $\mu_R'=\pm i\left(\frac{1}{2}-(\xi-i\chi)\right)$ and the state $\ket{0}_R$ with energy given by Eq.\eqref{0Reng} can likewise be constructed by adding left boundary string $\mu_R'=\pm i\left(\frac{1}{2}-(\xi-i\chi)\right)$ on top of all the real roots with respect to the reference state with all up spins. All the other excited states in this phase can be constructed by adding the bulk excitations (spinons, bulk strings, and quartets) on top of these elementary excitations. Since the bulk excitations have purely real energies, adding them on top of the state in $\mathscr{PT}$-symmetric phases results in states in excited states in $\mathscr{PT}$-symmetric phases and similarly adding these excitations on top of the state in spontaneously broken $\mathscr{PT}$ phase results in excited states in spontaneously broken $\mathscr{PT}$ phase as explained earlier in detail.

\subsubsection{{\color{blue}{$A_2 $ Phase}}}:
Consider the phase $A_2$ where the real part of the boundary parameter takes the value $0>\xi>-\frac{1}{2}$. In this phase, all the states can be constructed from the phase $A_1$ by using the property Eq.\eqref{symeqn}. 

We can take all the states in phase $A_1$ and apply the transformation that all the spins up and spins downs are interchanged and both the real and imaginary parts of the boundary parameters change the signs.  
\begin{equation}
\ket{\uparrow}\leftrightarrow \ket{\downarrow};\quad \xi\to -\xi \quad \text{and} \quad \chi\to -\chi
\label{prescriptiontt}
\end{equation}
to construct all the states in phase $A_2$

\subsection{B Phases }
\subsubsection{{\color{blue}{$B_1 $ Phase}}}:
Let us consider the parameter regime $\xi>\frac{1}{2}$.  The BAE with respect to the reference state with all spins up is 
\begin{equation}
\left(\frac{\mu_{j}-\frac{i}{2}}{\mu_{j}+\frac{i}{2}}\right)^{2 N} \left(\frac{\mu_{j}-i\left((\xi+i\chi)-\frac{1}{2}\right)}{\mu_{j}+i\left((\xi+i\chi)-\frac{1}{2}\right)}\right)\left(\frac{\mu_{j}-i\left((\xi-i\chi)-\frac{1}{2}\right)}{\mu_{j}+i\left((\xi-i\chi)-\frac{1}{2}\right)}\right) =\prod_{j \neq \ell=1}^{M}\left(\frac{\mu_{j}-\mu_{\ell}-i}{\mu_{j}-\mu_{\ell}+i}\right)\left(\frac{\mu_{j}+\mu_{\ell}-i}{\mu_{j}+\mu_{\ell}+i}\right),
\end{equation}
which can be written as
\begin{equation}
\left(\frac{\mu_{j}-\frac{i}{2}}{\mu_{j}+\frac{i}{2}}\right)^{2 N}= \left(\frac{\mu_j-\chi+i\left(\xi-\frac{1}{2}\right)}{\mu_j-\chi-i\left(\xi-\frac{1}{2}\right)} \right) \left(\frac{\mu_j+\chi+i\left(\xi-\frac{1}{2}\right)}{\mu_j+\chi-i\left(\xi-\frac{1}{2}\right)} \right) \prod_{j \neq \ell=1}^{M}\left(\frac{\mu_{j}-\mu_{\ell}-i}{\mu_{j}-\mu_{\ell}+i}\right)\left(\frac{\mu_{j}+\mu_{\ell}-i}{\mu_{j}+\mu_{\ell}+i}\right).
\end{equation}
The complex solution of the form $\mu_j=a+ib$ with $b>0$ makes the numerator of the LHS vanish, but the complex $\mu_j$ for which RHS vanishes are of the from $\mu_j=\pm\chi-i\left(\xi-\frac{1}{2}\right)$ \textit{i.e.} the imaginary part is negative. Thus, there are no boundary string solutions in this region.

\underline{\centering{$B_1$ Phase: Odd number of sites}}
The ground state is constructed from the reference state with all spins down. The root density is of the form
\begin{equation}
\tilde\rho_{\left|-\frac12\right\rangle}(\omega)=\frac{(2 N+1) e^{-\frac{| \omega | }{2}}-e^{- \left(\frac12+ \xi +i\chi\right) | \omega | }-e^{- \left( \frac12+\xi-i\chi\right) | \omega | }-1}{2 \left(e^{-| \omega | }+1\right)}.
\end{equation}
The total number of Bethe roots is given by
\begin{equation}
M_{\left|-\frac12\right\rangle}=\tilde\rho_{\left|-\frac12\right\rangle}(\omega=0)=\frac{N-1}{2}.
\end{equation}
We can compute the $z-$component of the total spin, where
\begin{equation}
S^z=-\left(\frac{N}{2}-M_{\left|-\frac12\right\rangle}\right)=-\frac{1}{2}.
\end{equation}

The energy of the state is $E=E_0$. 
All the other excited states in this phase can be constructed by adding the bulk excitations (spinons, bulk strings, and quartets) on top of the ground state. Since the ground state energy is real in this phase, all the excited states constructed by adding bulk excitations on top of this state have real energies. Hence all the states in this phase are in $\mathscr{PT}$ unbroken phase. 

\underline{\centering{$B_1$ Phase: Even number of sites}}\\
Starting from the reference state with all spins down, the total number of roots is integer only when the number of sites is odd. Thus, we need to add a hole and consider the state with root density

\begin{equation}
\tilde\rho_{{\left|-1\right\rangle}_\theta}(\omega)=\frac{(2 N+1) e^{-\frac{| \omega | }{2}}-e^{- \left(\frac12+ \xi +i\chi\right) | \omega | }-e^{- \left( \frac12+\xi-i\chi\right) | \omega | }-1}{2 \left(e^{-| \omega | }+1\right)}+\Delta\tilde\rho_\theta(\omega).
\end{equation}
Now, the total number of roots is given by
\begin{equation}
M_{{\left|-1\right\rangle}_\theta}=\tilde\rho_{{\left|-1\right\rangle}_\theta}=\frac{N-2}{2}.
\end{equation}
Thus, the $z-$component of the spin is
\begin{equation}
S^z_{{\left|-1\right\rangle}_\theta}=-\left(\frac{N}{2}-\frac{N-2}{2}\right)=-1.
\end{equation}
The energy of this state is
\begin{equation}
E_{{\left|-1\right\rangle}_\theta}=E_0+\frac{2\pi}{\cosh(\pi\theta)}.
\end{equation}

We can create another state $\ket{0}_\theta$ by adding a spinon on top of the state with all real roots with respect to the reference state with all spins up. The root density becomes
\begin{equation}
\tilde\rho_{{\left|0\right\rangle}_\theta}(\omega)=\frac{(2 N+1) e^{-\frac{| \omega | }{2}}+e^{- \left( \xi +i\chi-\frac12\right) | \omega | }+e^{- \left( \xi-i\chi-\frac12\right) | \omega | }-1}{2 \left(e^{-| \omega | }+1\right)}+\Delta\tilde\rho_\theta(\omega).
\end{equation}

The total number of roots is given by
\begin{equation}
M_{{\left|0\right\rangle}_\theta}=\rho_{{\left|0\right\rangle}_\theta}(0)=\frac{N}{2},
\end{equation}
and hence the $z-$component of the spin is
\begin{equation}
S_{{\left|0\right\rangle}_\theta}^z=0.
\end{equation}
Once again using Eq.\eqref{engcont}, the energy of the state is obtained as
\begin{equation}
E_{{\left|0\right\rangle}_\theta}=E_0+\frac{2\pi}{\cosh(\pi\theta)}.
\end{equation}
The minimal energy state is obtained in the limit $\theta\to\infty$. Thus, the two-fold degenerate ground state is represented by $\ket{0}_{\theta\to\infty}$ and $\ket{-1}_{\theta\to\infty}$.

\subsubsection{{\color{blue}{$B_2 $ Phase}}}

Let us consider the parameter regime $\xi<-\frac{1}{2}$. 

\underline{\centering{$B_2$ Phase: Odd number of sites}}\\
The ground state is constructed from the reference state with all spins up. The root density is of the form
\begin{equation}
\tilde\rho_{\left|\frac12\right\rangle}(\omega)=\frac{(2 N+1) e^{-\frac{| \omega | }{2}}-e^{- \left(\frac12- \xi -i\chi\right) | \omega | }-e^{- \left( \frac12-\xi+i\chi\right) | \omega | }-1}{2 \left(e^{-| \omega | }+1\right)}.
\end{equation}
The total number of Bethe roots is given by
\begin{equation}
M_{\left|\frac12\right\rangle}=\tilde\rho_{\left|\frac12\right\rangle}(\omega=0)=\frac{N-1}{2},
\end{equation}
from which we can compute the $z-$component of the total spin, where
\begin{equation}
S^z=\left(\frac{N}{2}-M_{\left|\frac12\right\rangle}\right)=\frac{1}{2}.
\end{equation}

The energy of the state is $E=E_0$. All the other excited states in this phase can be constructed by adding the bulk excitations (spinons, bulk strings, and quartets) on top of these doubly degenerate ground states. Since both the ground states have completely real energy in this phase, all the excited states constructed by adding bulk excitations on top of these states have real energies. Hence all the states in this phase are in $\mathscr{PT}$ unbroken phase.

\underline{\centering{$B_2$ Phase: Even number of sites}}\\
Using the transformation Eq.\eqref{prescriptiontt}, we can construct the state in this phase from the $B_1$ phase with an even number of sites. Here, the ground state is two-fold degenerate $\ket{0}_{\theta\to\infty}$ and $\ket{1}_{\theta\to\infty}$. The former state is created by adding a spinon on the top of the real roots with respect to the reference state with all spins down, and the latter is created by adding a spinon on the top of the real roots with respect to the reference state with all spins up. The former has $\frac{N}{2}$ number of  roots and $S^z=0$, and the latter has $\frac{N-2}{2}$ number of the real roots and spin $S^z=1$.

\section{Wavefunction in one particle sector}
\label{sec:WF}
Following \cite{alcaraz1987surface, skorik1995boundary,XXXmag}, we study the pure boundary effect by computing the wavefunction in the 1-magnon sector.

The Bethe Ansatz equation in the 1-magnon sector becomes
\begin{equation}
     \left(\frac{\mu-\frac{i}{2}}{\mu+\frac{i}{2}}\right)^{2N}\frac{\mu+i\left(\frac{1}{2}-\xi_-\right)}{\mu-i\left(\frac{1}{2}-\xi_-\right)}\frac{\mu+i\left(\frac{1}{2}-\xi_+\right)}{\mu-i\left(\frac{1}{2}-\xi_+\right)}=1.
\end{equation}
And the wavefunction reads
\begin{equation}
    F(x)=\left(1+\frac{\left(\mu -\frac{i}{2}\right) }{\mu +\frac{i}{2}}\left(\frac{1}{\xi _-}-1\right)\right) \left(\frac{\mu +\frac{i}{2}}{\mu -\frac{i}{2}}\right)^x-\left(1+\frac{\left(\mu +\frac{i}{2}\right) }{\mu -\frac{i}{2}}\left(\frac{1}{\xi _-}-1\right)\right) \left(\frac{\mu +\frac{i}{2}}{\mu -\frac{i}{2}}\right)^{-x}
    \end{equation}

For $\mu=\pm i(\frac{1}{2}-\xi_-)$, the above wavefunction reduces to the non-normalized wavefunction of the left bound mode
\begin{equation}
    F_L(x)=\pm\frac{\left(2 \xi _--1\right) }{\xi _-^2}\left(\frac{\xi _--1}{\xi _-}\right)^{-x}.
   \label{lsolpsi}.
\end{equation}

For $\mu=\pm i(\frac{1}{2}+\xi_-)$, we get the non-normalized wavefunction of the right bound mode
\begin{equation}
   F_R(x)=\pm\frac{1-2 \xi _+}{\left(\xi _+-1\right) \xi _+}\left(\frac{\xi _+-1}{\xi _+}\right)^{-(N-x)}.
   \label{rsolpsi}
\end{equation}

\begin{figure}[H]
\centering
\begin{subfigure}[t]{0.4\linewidth}
    \centering
    \includegraphics[width=\linewidth]{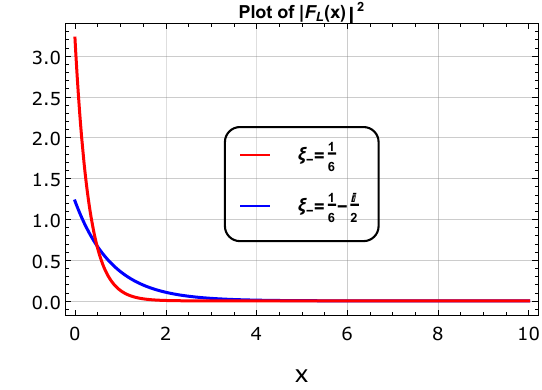}
    \caption{Modulus of the square of the wavefunction for the left localized bound mode.}
    \label{fig:wfnl}
\end{subfigure}\hfil
\begin{subfigure}[t]{0.4\linewidth}
    \centering
    \includegraphics[width=\linewidth]{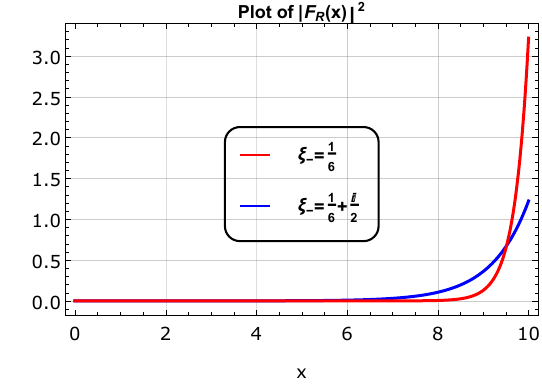}
    \caption[b]{Modulus of the square of the wavefunction for the right localized bound mode.}
     \label{fig:wfnr}
 \end{subfigure}
 \caption{Plots of the modulus of squares of wavefunctions for $N=10$.}
 \label{fig:wfn}
\end{figure}

Properly normalizing the modulus of the squares of wavefunctions given by Eq.\eqref{lsolpsi} and Eq.\eqref{rsolpsi}, we can plot the modulus squared wavefunction for both real and complex boundary fields. As shown in Figure \ref{fig:wfn}, we see that the wavefunction is exponentially localized sharply at the boundary in the case of the real boundary fields and as the imaginary part of the field is increased, the localized wavefunction starts to broaden thereby making the demarcation of the bulk and boundary flimsy.

\section{Summary and Discussion}
\label{sec:discussion}
We considered the integrable Heisenberg spin chain with complex boundary fields and diagonalized by using the Bethe Ansatz method. For generic complex boundary fields, the roots of the Bethe equations lie in a complex plane and it is not, in general,  possible to solve the resulting integral equation. However, for the choice of $\mathscr{PT}$ - symmetric boundary fields, the roots of Bethe equations lie on the real line which enables us to solve the resulting integral equation for the density of the roots. This choice corresponds to the case where there is a balanced loss and gain in the model. For this choice of the boundary fields, we have shown that the system exhibits two types of phases named $A$ and $B$. In the $B$ type phase all the eigenstates have real energies in the thermodynamic limit, hence the $\mathscr{PT}$ symmetry is unbroken. The $A$ type phase exhibits two sectors, where one sector comprises of eigenstates with real energies and the other sector comprises of eigenstates with complex energies, corresponding to $\mathscr{PT}$ symmetry unbroken and broken sectors respectively. Furthermore, we find that the ground state exhibited by the system changes depending on the orientation of the boundary fields. Each of the $A$ and $B$ type phases can be divided into two sub-phases named $A_1$, $A_2$ and $B_1$, $B_2$ respectively, depending on the ground state exhibited by the system. 

The existence of two sectors in the $A$ type phase is related to the existence of localized bound states at the left and right edges, whose energies are complex conjugates of each other. We computed the wavefunction in one magnon sector and we find that the exponentially localized boundary wavefunction starts to broaden as we increase the complex part of the boundary field. In addition to this, as the complex part of the boundary fields is increased, the energy of the bound state becomes lesser than the maximum energy of a single spinon which is $2\pi$. Taking into account the broadening of the bound state wavefunction, and the merging of the energy of the bound state into the band of a single spinon branch, it is natural to interpret that for large values of $\chi$ the boundary string no longer corresponds to a bound state exponentially localized at the edge.

Recently, it was shown \cite{XXXmag} that in the Heisenberg spin chain with real boundary fields, the Hilbert space comprises of a certain number of towers which depends on the number of bound states exhibited by the system. It was also shown that as the boundary parameters are changed, the bound states leak into the bulk, which results in an eigenstate phase transition where the number of towers of the Hilbert space changes. It is natural that the Heisenberg chain with complex boundary fields that we considered here may also exhibit such towers of excited states in the $A$ type phases, where there exists exponentially localized bound states at the edges. If so, it would be interesting to understand the fate of these towers when one moves from $A$ type to $B$ type phases and also for large values of $\chi$ where the bound states may no longer exist. 

\section*{References}
\bibliographystyle{unsrt}
\bibliography{ref}
\end{document}